\documentclass[journal]{IEEEtran}
\IEEEoverridecommandlockouts
\usepackage{type1ec}
\usepackage[T1]{fontenc}
\usepackage{cite}
\usepackage{amsmath,amssymb} % comment symbol disable

\usepackage{algorithmic}
\usepackage{float} % for floating figures
\usepackage{pstricks}
\usepackage{pstricks-add}
\usepackage{pst-circ}
\usepackage{subcaption}
\usepackage{graphicx}
\usepackage{textcomp}
\usepackage{xcolor}
\usepackage{pst-coil}
\usepackage{pst-node}
\usepackage{pst-all}	%call the pstricks package
\usepackage{pst-func}
\usepackage{array}
\usepackage{placeins}
\usepackage{siunitx} 

\def\BibTeX{{\rm B\kern-.05em{\sc i\kern-.025em b}\kern-.08em
    T\kern-.1667em\lower.7ex\hbox{E}\kern-.125emX}}
\usepackage[ruled,vlined]{algorithm2e}
\usepackage{slashbox}
\usepackage{mathtools}

\usepackage{multirow}
\usepackage[bookmarks=false]{hyperref}
\usepackage{url}
\begin{document}
\begin{NoHyper}

\title{Terabit-per-Second Multicore Polar Code Successive Cancellation Decoders\\
}

% \author{\IEEEauthorblockN{Altu\u{g}~S\"{u}ral,  Ertu\u{g}rul  Kola\u{g}as{\i}o\u{g}lu and Erdal~Ar{\i}kan}
%\IEEEauthorblockA{Department of Electrical and Electronics Engineering, Bilkent University, Turkey \\
%\{altug, arikan\}@ee.bilkent.edu.tr } 
% }

% $^3$Department of Electrical and Electronics Engineering, Bilkent University, Turkey \\
%  and Erdal~Ar{\i}kan and E.~Ar{\i}kan
%  and Erdal~Ar{\i}kan
\author{Altu\u{g}~S\"{u}ral, Ertu\u{g}rul  Kola\u{g}as{\i}o\u{g}lu% <-this % stops a space
\thanks{Altu\u{g} S\"{u}ral, Ertu\u{g}rul  Kola\u{g}as{\i}o\u{g}lu are with POLARAN, Ankara, Turkey (e-mails: \{altug.sural, ekolagasioglu\}@polaran.com).}% <-this % stops a space
\thanks{A. S\"{u}ral is also with Department of Electrical and Electronics Engineering, Bilkent University, Turkey  (e-mail: altug@ee.bilkent.edu.tr).}% <-this % stops a space
\thanks{Manuscript received October 10, 2020}} % ; revised August 26, 2015.
%With this method, the pipeline depth is kept at 60 clock cycles by merging consecutive decoding stages. 

\maketitle
\begin{abstract}
This work presents a high throughput and energy efficient multicore (MC) successive cancellation (SC) decoder architecture for polar codes. SC is a low-complexity decoding algorithm with a set of sequential operations. The sequential processing nature of SC limits parallelism but promotes not only pipelining but also multiple copies of SC decoder with an optimized pipeline depth to achieve Tb/s throughput. The MCSC decoder architecture consists of multiple SC decoders with lower frequency and pipeline depth to process multiple codewords in parallel to achieve lower power consumption. The pipeline depth of MCSC is optimized separately for each multicore configuration using register reduction/balancing (R-RB) method. This enables an efficient implementation for the 1-core, 2-core 4-core and 8-core candidate MCSC decoders. To reduce the complexity of the implementation, an adaptive log-likelihood ratio (LLR) quantization scheme is used for internal LLRs within the range of 1-5 bits. The post-placement-routing results at 28nm High-k Metal Gate (HKMG) ASIC technology show that 4-core MCSC decoder achieves 1 Tb/s throughput on 3.92 mm$^2$ area with 1.55 pJ/bit energy efficiency.
\end{abstract}

% ----------------------------------------
\section{Introduction} \label{sec:intro}
% ----------------------------------------
%%% rewrite
Polar codes can achieve channel capacity in broad class of channels using a low-complexity successive cancellation (SC) decoder \cite{Arikan2009}. SC is a depth-first search decoding algorithm, which tracks the best known decoding path in a sequential manner. SC has a mediocre error correction performance at moderate data block lengths. An important reason is that it does not update any given bit estimate according to later redundancy information. SC-list (SCL) decoding algorithm is proposed to improve decoding performance of the SC up to maximum-likelihood decoding performance \cite{Tal2015}. SCL algorithm tracks a list of possible decoding paths and reaches near ML performance with an additional computational complexity. The performance of SC algorithm can also be improved by enabling multiple iterations that allow updating bit estimates. SC-flip (SCF) \cite{Afisiadis2014} and Soft-cancellation (SCAN) \cite{Fayyaz2014} algorithms utilize multiple iterations to fix decision errors of SC. 

Due to inherently sequential structure of SC, advanced hardware architectures are required to achieve high throughput. Available pipelined SC decoder implementations in the literature \cite{Zhang2013}, \cite{Dizdar2016}, \cite{Giard2015}, \cite{Giard2016}  can only achieve on the order of Gb/s throughput. The unrolled implementations in \cite{Giard2015}, \cite{Giard2016} exploit shortcuts introduced in \cite{Yazdi2011}, \cite{Sarkis2014}, \cite{Hanif2017} to improve hardware efficiency. For  higher throughput, quantization of soft-information and register reduction/balancing methods have been proposed in \cite{Sural2019}. A generic problem identified for Tb/s throughput regime is power density caused by excessive switching activity in a limited core area. The power density can be reduced by standard IC design techniques such as using a low-power cells having a reduced supply voltage.

In this paper, an architectural method is proposed to reduce power dissipation using multicore SC decoders. The MCSC decoder architecture consists of multiple SC decoders with lower frequency and optimized pipeline depth to process multiple codewords in parallel. The pipeline depth of MCSC is optimized separately for each multicore configuration using register reduction/balancing (R-RB) method. An adaptive log-likelihood ratio (LLR) quantization scheme is used to reduce implementation complexity. These methods enable an efficient implementation for the 1-core, 2-core 4-core and 8-core candidate MCSC decoders.
% ----------------------------------------
\subsection{Main contributions}  \label{sec:achievements}
% ----------------------------------------
The main contributions of this paper are as follows.
\begin{itemize}
\item Multicore VLSI architecture has not been proposed and investigated with fair comparison for high throughput polar code decoders. Full timing clean post place and route results are given for 1-core, 2-core, 4-core and 8-core SC decoder designs. The proposed MCSC decoder is the first multicore polar decoder that achieves 1 Tb/s throughput.
\item MCSC decoder and a channel simulator (capable of simulating very low error rates) are implemented on FPGA to verify RTL code and measure error correction performance at very low BER. The FPGA implementation results show that MCSC decoder achieves 200 Gb/s throughput and $1.1 \times 10^{-13}$ bit error rate (BER) at 8 dB Eb/No.
\item In addition to its multi-core architecture, MCSC decoder exploits shortcuts, adaptive quantization and R-RB methods to reduce design area and power.
\item The results also show that 4-core MCSC decoder archives 1.55 pJ/bit energy efficiency at 3.92 mm$^2$ area. This energy efficiency is significantly lower than any reported very high throughput FEC decoder.
\end{itemize}

The outline of this paper is as follows. Section \ref{sec:preliminaries} gives a short summary of polar coding and SC decoding algorithm. Section \ref{sec:proposed_multicore} presents the proposed MCSC decoder architecture including register reduction/balancing and adaptive quantization. Section \ref{sec:fpga_verification} presents FPGA verification and communication performance of the proposed decoder. Section \ref{sec:asic_implemetation} presents ASIC implementation details and show comparison with state-of-the-art implementations. Finally, Section \ref{sec:conclusion} summarizes the main results with a brief conclusion.
% ----------------------------------------
\section{Preliminaries}  \label{sec:preliminaries}
% ----------------------------------------
% ----------------------------------------
\subsection{Polar codes}   \label{sec:polar_code_review}
% ----------------------------------------
% rewrite
Polar codes are linear block codes that have a polar transform matrix $G_N = G^{\otimes n}$ with a code block length $N=2^n$ for each $n\ge 1$. The polar transform matrix can be recursively calculated with the $n^{\text{th}}$ Kronecker power of the generator matrix $G =\begin{bmatrix}
1 & 0 \\
1 & 1 
\end{bmatrix}$. 
Polar codes provides a different degree of protection for the data bits at channel indices due to the polarization effect. The most reliable channel indices constitute a information (free) data set $u_{\cal A}$, where length-$K$ user data $d_K$ is assigned. The remaining indices constitute a frozen data set $u_{{\cal A}^c}$, where a set of predetermined values (typically zero) are assigned. A collection of free and frozen bits forms an input transform vector $u_N$. Polar encoding operation transforms $u_N$ to a valid codeword $x_N$ with the matrix multiplication $x_N = u_N G_N$. 

% ----------------------------------------
\subsection{SC Decoding Algorithm}  \label{sec:sc_algorithm}
% ----------------------------------------
SC decoding algorithm is given in Algorithm \ref{sc_algorithm}. The algorithm takes LLR vector $\ell_M$, indicator vector of frozen bits $v_M$ and recursive block length parameter $M$. Initially, $\ell_M = \frac{2y_i}{\sigma^2}$ for an AWGN channel $W$ with the variance $\sigma ^2$ and $M=N$. The $v_M$ vector consists of one values for the frozen indices and zero values for the free indices of the target polar code. For example, $v_M = \{1,0\}$ for the ($N=2$, $K=1$) polar code. The input values are updated recursively for the first and second recursions of the algorithm. The LLR input of the first and second recursions is calculated by F and G functions, respectively. These functions decompose a length-$M$ polar code into two length-$M/2$ polar code segments and our implementations are explicitly defined in \cite{Sural2019}. The G functions require a hard decision feedback $\hat{z}_{M/2}$, which can be calculated after a set of frozen and free hard decisions.

At the end of final recursion, an estimate of the user data $\hat{d}_K$, belongs to $u_{\cal A}$, is extracted from an estimate of the input transform vector $\hat{u}_N$. We refer to \cite{Arikan2009} for a complete description of the SC decoding algorithm.

\SetKwData{Left}{left}\SetKwData{This}{this}\SetKwData{Up}{up}
\SetKwFunction{Union}{Union}\SetKwFunction{FindCompress}{FindCompress}
\SetKwInOut{Input}{input}\SetKwInOut{Output}{output}
\SetArgSty{textnormal}
\begin{algorithm}
\renewcommand{\algorithmicrequire}{\textbf{Inputs   :}}
\renewcommand{\algorithmicensure}{\textbf{Output:}}
\algorithmicrequire{ $\ell_M$, $v_M$, $M$} \text{  } \algorithmicensure{ $\hat{u}_M$}\\
  \uIf(\tcp*[f]{\small frozen}){$v_M= 1$ \bf{and} $M = 1$}{
  $\hat{u}_M  = 0$\\
  }
  \uElseIf(\tcp*[f]{\small free}){$v_M=0$ \bf{and} $M = 1$}{
  $\hat{u}_M =\text{d}(\ell_M$, $v_M= 0)$  \\ 
	  \uIf(\tcp*[f]{\small positive LLR value}){$\ell_M \geq 0$ }{
		  $\hat{u}_M  = 0$\\
	  }
	  \Else(\tcp*[f]{\small negative LLR value}){ 
		$\hat{u}_M  = 1$\\
      }
  }
  \Else(\tcp*[f]{\small Decode two length-$M/2$ polar codes}){ 
   $l_{M/2} = \text{F}(\ell_M^{\text{odd}}$, $\ell_M^{\text{even}}$)  \tcp*[f]{\small F function} \\
   $\hat{z}_{M/2} = \text{SC}(l_{M/2}$, $v_M^{\text{odd}}$, $\frac{M}{2})$ \tcp*[f]{\small First recursion} \\
   $r_{M/2} = \text{G}(\ell_M^{\text{odd}}$, $\ell_M^{\text{even}}$, $\hat{z}_{M/2})$ \tcp*[f]{\small G function} \\
   $\hat{x}_{M/2} = \text{SC}(r_{M/2}$, $v_M^{\text{even}}$, $\frac{M}{2}$)\tcp*[f]{\small Second recursion} \\
   $\hat{u}_M^{\text{odd}} = \hat{z}_{M/2}$ $\oplus$ $\hat{x}_{M/2}$ \tcp*[f]{\small Calculate feedback}\\
   $\hat{u}_M^{\text{even}} = \hat{x}_{M/2}$\\
   }
   \uIf(\tcp*[f]{\small User data extraction}){$M = N$ }{
	 $\hat{d}_K = u_{\cal A}$ \\
	 \Return {$\hat{d}_K$}
   }
   \Else(\tcp*[f]{\small Decode remaining code segments}){ 
     \Return {$\hat{u}_M$}
   }
\caption{SC}
\label{sc_algorithm}
\end{algorithm}

% ----------------------------------------
\subsubsection{Shortcuts for making multiple parallel decisions} \label{sec:shortcuts}
% ----------------------------------------.
In SC algorithm, as larger code segments are divided into smaller ones some special code structures are encountered where multiple decisions can be calculated in parallel using certain shortcuts introduced in \cite{Sarkis2014}. These easily decodable polar code segments consist of single-parity-check, repetition, rate-1 and rate-0 codes. We also take advantage of these shortcuts to increase the throughput of the SC decoder. Length of these shortcut blocks are also optimized in our implementation.
% ----------------------------------------
\section{Proposed Multicore Polar SC Decoder Architecture} \label{sec:proposed_multicore}
% ----------------------------------------

Initially, a single core SC decoder was investigated and synthesized to achieve 1 Tb/s data throughput. However, power density of this single core architecture was too high due to large number of pipeline stages that have to be clocked with a high frequency. To improve the design, multicore architecture is evaluated. In this architecture, each decoder core has unrolled and pipelined hardware architecture to achieve Tb/s throughput. Main motivations for our multicore SC decoder approach can be summarized as follows.
\begin{itemize}
\item Even for the same pipeline depth, the multicore architecture does not have any power penalty. Thanks to reduced clock frequency, the total data movement is the same at worst.
\item With reduced clock frequency of cores, it is possible to reduce pipeline depth. This not only reduces total data movement but also minimizes buffer sizes of the  recursive SC decoder. As a result, significant power savings with minimal area cost is obtained.
\item It is possible to manage power density by arranging core positions in floorplan of hierarchical design.
\item Better combinational logic optimization and better clock period utilization are also expected.
\item Lower frequency of operation will minimize metal migration problem.
\end{itemize}
To make a fair comparison same ($R=5/6$) code rate and same throughput ($\gamma=1$ Tb/s) throughput is used for all designs. Number of cores ($P$) and core clock frequency ($f_c$) pairs are (1, 1200 MHz), (2, 600 MHz), (4, 300 MHz), (8, 150 MHz). The multicore polar SC decoder architecture is shown in Figure \ref{fig:multi_core_architecture}. 
\begin{figure}[ht]
	\centering
	\includegraphics[scale=0.43]{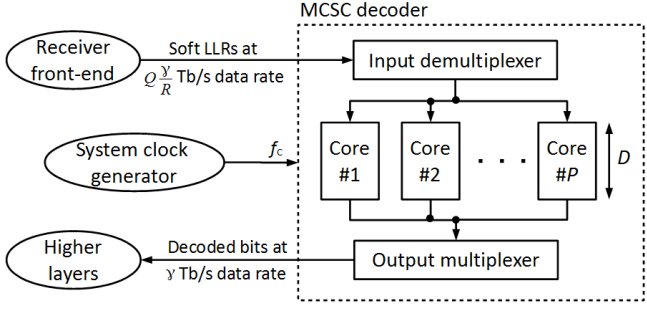}
	\caption{Multicore SC decoder architecture}
	\label{fig:multi_core_architecture}
\end{figure}
Another very important aspect of fair comparison between multicore designs is number of IO pins. For the single core to achieve 1 Tb/s throughput, we need 5120 input (5b soft inputs) and 854 output pins operating at 1200MHz. We have decided that we should not increase the number of IO pins any further as we increase the number of cores ($P$). As a result, input and output multiplexers are introduced in the multicore architecture to distribute the soft information and collect estimated information bits to and from the decoder cores. The input output clock frequency $f_{IO}$ is equal to $P f_c$ to provide sufficiently high throughput at the interface. This is also a balancing cost factor while we are increasing number of cores.

The input demultiplexer and output multiplexer can also be placed inside the decoder cores to reduce the complexity of data distribution network and clock domain crossing overhead as shown in Fig. \ref{fig:multicore_distributed}. Each core has $\frac{NQ}{P}$ number of LLR input pins and the LLR inputs are stored at the input of each core using a shift register with $NQ$ depth. Similarly, the outputs are stored at the final pipeline stage before given to higher layers. This architecture option is selected in this work to reduce implementation complexity and also power consumption. 

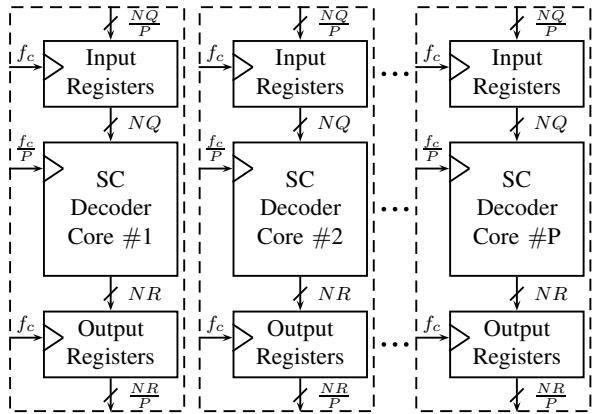
\begin{figure} % [ht]
\centering
\begin{pspicture}(0,0)(8.5,5.4)
%\psgrid[gridcolor=green,subgridcolor=yellow]
\psscalebox{0.9}{
\rput(0.5,-0.5){
\rput(-0.5,0){
\psframe(1,5)(3,6) \rput(2,5.5){\begin{tabular}{c} Input\\ Registers \end{tabular}}
\psframe(1,2.5)(3,4.5) \rput(2,3.5){\begin{tabular}{c} SC\\Decoder\\ Core $\#$1 \end{tabular}}
\psframe(1,1)(3,2) \rput(2,1.5){\begin{tabular}{c} Output\\ Registers \end{tabular}}
\psline{->}(2,5)(2,4.5) \psline(1.9,4.65)(2.1,4.85) \rput(2.5,4.75){\fontsize{8pt}{8pt}$NQ$} % \text{Decoder}
\psline{->}(2,2.5)(2,2) \psline(1.9,2.15)(2.1,2.35) \rput(2.5,2.25){\fontsize{8pt}{8pt}$NR$} 
\psline{->}(2,6.5)(2,6) \psline(1.9,6.15)(2.1,6.35) \rput(2.5,6.25){\fontsize{8pt}{8pt}$\frac{NQ}{P}$} 
\psline{->}(2,1)(2,0.5) \psline(1.9,0.65)(2.1,0.85) \rput(2.5,0.75){\fontsize{8pt}{8pt}$\frac{NR}{P}$} 
% triangulars
\psline(1,1.8)(1.3,1.6) \psline(1.3,1.6)(1,1.4) \psline{->}(0.5,1.6)(1,1.6) \rput(0.75,1.8){\fontsize{8pt}{8pt}$f_{c}$} 
\rput(0,2.5){\psline(1,1.8)(1.3,1.6) \psline(1.3,1.6)(1,1.4) \psline{->}(0.5,1.6)(1,1.6) \rput(0.75,1.9){\fontsize{8pt}{8pt}$\frac{f_{c}}{P}$} }
\rput(0,4){\psline(1,1.8)(1.3,1.6) \psline(1.3,1.6)(1,1.4) \psline{->}(0.5,1.6)(1,1.6) \rput(0.75,1.8){\fontsize{8pt}{8pt}$f_{c}$} }
% top frame
\psframe[linestyle=dashed](0.5,0.5)(3.1,6.5)}

\rput(2.3,0){
\psframe(1,5)(3,6) \rput(2,5.5){\begin{tabular}{c} Input\\ Registers \end{tabular}}
\psframe(1,2.5)(3,4.5) \rput(2,3.5){\begin{tabular}{c} SC\\Decoder\\ Core $\#$2 \end{tabular}}
\psframe(1,1)(3,2) \rput(2,1.5){\begin{tabular}{c} Output\\ Registers \end{tabular}}
\psline{->}(2,5)(2,4.5) \psline(1.9,4.65)(2.1,4.85) \rput(2.5,4.75){\fontsize{8pt}{8pt}$NQ$} % \text{Decoder}
\psline{->}(2,2.5)(2,2) \psline(1.9,2.15)(2.1,2.35) \rput(2.5,2.25){\fontsize{8pt}{8pt}$NR$} 
\psline{->}(2,6.5)(2,6) \psline(1.9,6.15)(2.1,6.35) \rput(2.5,6.25){\fontsize{8pt}{8pt}$\frac{NQ}{P}$} 
\psline{->}(2,1)(2,0.5) \psline(1.9,0.65)(2.1,0.85) \rput(2.5,0.75){\fontsize{8pt}{8pt}$\frac{NR}{P}$} 
% triangulars
\psline(1,1.8)(1.3,1.6) \psline(1.3,1.6)(1,1.4) \psline{->}(0.5,1.6)(1,1.6) \rput(0.75,1.8){\fontsize{8pt}{8pt}$f_{c}$} 
\rput(0,2.5){\psline(1,1.8)(1.3,1.6) \psline(1.3,1.6)(1,1.4) \psline{->}(0.5,1.6)(1,1.6) \rput(0.75,1.9){\fontsize{8pt}{8pt}$\frac{f_{c}}{P}$} }
\rput(0,4){\psline(1,1.8)(1.3,1.6) \psline(1.3,1.6)(1,1.4) \psline{->}(0.5,1.6)(1,1.6) \rput(0.75,1.8){\fontsize{8pt}{8pt}$f_{c}$} }
% top frame
\psframe[linestyle=dashed](0.5,0.5)(3.1,6.5)}

\rput(5.5,0){
\psframe(1,5)(3,6) \rput(2,5.5){\begin{tabular}{c} Input\\ Registers \end{tabular}}
\psframe(1,2.5)(3,4.5) \rput(2,3.5){\begin{tabular}{c} SC\\Decoder\\ Core $\#$P \end{tabular}}
\psframe(1,1)(3,2) \rput(2,1.5){\begin{tabular}{c} Output\\ Registers \end{tabular}}
\psline{->}(2,5)(2,4.5) \psline(1.9,4.65)(2.1,4.85) \rput(2.5,4.75){\fontsize{8pt}{8pt}$NQ$} % \text{Decoder}
\psline{->}(2,2.5)(2,2) \psline(1.9,2.15)(2.1,2.35) \rput(2.5,2.25){\fontsize{8pt}{8pt}$NR$} 
\psline{->}(2,6.5)(2,6) \psline(1.9,6.15)(2.1,6.35) \rput(2.5,6.25){\fontsize{8pt}{8pt}$\frac{NQ}{P}$} 
\psline{->}(2,1)(2,0.5) \psline(1.9,0.65)(2.1,0.85) \rput(2.5,0.75){\fontsize{8pt}{8pt}$\frac{NR}{P}$} 
% triangulars
\psline(1,1.8)(1.3,1.6) \psline(1.3,1.6)(1,1.4) \psline{->}(0.5,1.6)(1,1.6) \rput(0.75,1.8){\fontsize{8pt}{8pt}$f_{c}$} 
\rput(0,2.5){\psline(1,1.8)(1.3,1.6) \psline(1.3,1.6)(1,1.4) \psline{->}(0.5,1.6)(1,1.6) \rput(0.75,1.9){\fontsize{8pt}{8pt}$\frac{f_{c}}{P}$} }
\rput(0,4){\psline(1,1.8)(1.3,1.6) \psline(1.3,1.6)(1,1.4) \psline{->}(0.5,1.6)(1,1.6) \rput(0.75,1.8){\fontsize{8pt}{8pt}$f_{c}$} }
% top frame
\psframe[linestyle=dashed](0.5,0.5)(3.1,6.5)}

\psline[linestyle=dotted,linewidth=0.07](5.5,3.5)(5.9,3.5)
\psline[linestyle=dotted,linewidth=0.07](5.5,5.5)(5.9,5.5)
\psline[linestyle=dotted,linewidth=0.07](5.5,1.5)(5.9,1.5)
} }      
\end{pspicture}
\caption{Revised multicore SC decoder architecture}
\label{fig:multicore_distributed}
\end{figure}

% ----------------------------------------
\subsection{Adaptive quantization of the LLRs} \label{sec:adapt_quantization}
% ----------------------------------------
Adaptive quantization is an optimization scheme to reduce storage and computational complexity by decreasing LLR bit precision of internal data paths in the SC decoder. Due to polarization during SC decoding, using constant number of LLR quantization bits is inefficient. As polarization increases reliability, LLR resolution can be decreased without significant performance loss. Adaptive quantization enables each internal LLR value to be represented by a variable number of bits. We refer to \cite{Sural2019} for the comprehensive definition of adaptive quantization scheme.

% ----------------------------------------
\subsection{Register reduction/balancing} \label{sec:register_balancing}
% ----------------------------------------
The pipeline buffers may consist of a large load for the clock network that can significantly increase the overall power consumption. Hence for efficient hardware implementation reducing number of pipeline stages is very critical optimization step. In order to reduce and optimize registers in SC decoder design, the R-RB technique has been performed. Together with the multicore architecture, pipeline depth of SC decoder core is reduced to 13 (for 8-core design) by merging the consecutive stages and removing the registers. Each pipeline stage has almost equal cell delay as a result of register balancing. As a result of bit reversed architecture and smooth data flow of SC decoder, total delay is assumed to be proportional to cell delays for decoder cores.
% ----------------------------------------
\subsection{Latency analysis of the proposed MCSC architecture} \label{sec:comp_analysis}
% ----------------------------------------
Let $D$ is the pipeline depth, $P$ is the number of decoder cores, $\theta$ is the phase of the internal clock $f_c$ in degrees and $T_{\text{IO}}$ is the period of the interface clock $f_{\text{IO}}$, then the latency of the multicore SC decoder architecture is 
\begin{align}
L = T_{\text{IO}}\bigg(P(D+2)+\text{mod}\left( \frac{P(\theta +180)}{360},P \right)\bigg)\text{.} \nonumber
\end{align}
The 4-core architecture with $P=4$, $D=25$ and $T_{\text{IO}}=0.833\text{ ns}$ has $ 89.9\text{ ns} \leq L \leq 92.5 \text{ ns}$ latency. The 8-core architecture with $P=8$, $D=12$ and $T_{\text{IO}}=0.833\text{ ns}$ has $ 93.2\text{ ns} \leq L \leq 99.2\text{ ns}$ latency depending on the phase of the internal clock.

% ---------------------------------------- 
\section{FPGA Verification and Performance} \label{sec:fpga_verification}
% ----------------------------------------
 
% ----------------------------------------
\subsection{FPGA test platform} \label{sec:awgn_fpga}
% ----------------------------------------
MCSC polar decoder implementations have been verified via an FPGA test platform shown in Fig. \ref{fig:fpga_test}. The test platform can support information throughput up to 200 Gb/s. For each transmitted polar codeword, a linear feedback shift register (LFSR) generates 854 bit pseudo random data. A systematic polar encoder generates 1024 bit encoded data from the pseudo random data. The encoded data consists of 854 systematic bits and 170 parity bits, where both bits are mapped to BPSK symbols. The symbols are accumulated with AWGN generated by a build-in Gaussian random number generator. After that a fixed point sign-magnitude LLR values are obtained from the output of the AWGN channel. The polar SC decoder gets LLRs and produces information bit estimates, which are compared with the original data to achieve error statistics. 

\begin{figure}
	\centering
	\includegraphics[scale=0.35]{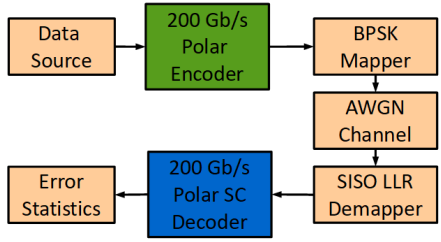}
	\caption{Block diagram of FPGA test platform}
	\label{fig:fpga_test}
\end{figure}

% ----------------------------------------
\subsection{FPGA performance results} \label{sec:fpga_performance}
% ----------------------------------------
The frame error performance results in Fig. \ref{fig:sc_fer} show that adaptively quantized FPGA implementation of MCSC decoder causes approximately 0.25 dB Eb/No performance loss at 10$^{-5}$ FER compared to floating-point software simulation. The 0.1 dB performance loss is caused by adaptive quantization of LLRs on FPGA.  
\begin{figure}[ht]
	\centering
	\includegraphics[scale=0.6]{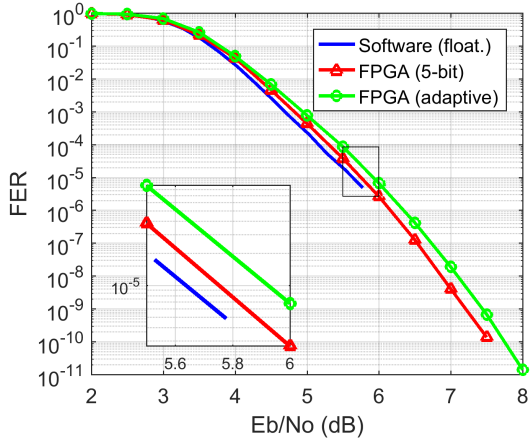}
	\caption{FER performance of (1024,854) polar code with SC decoder}
	\label{fig:sc_fer}
\end{figure}
Moreover, BER performance results in Fig. \ref{fig:sc_ber} show that a coding gain of 6.21 dB is attained at $10^{-12}$ BER relative to uncoded transmission.
\begin{figure}[ht]
	\centering
	\includegraphics[scale=0.6]{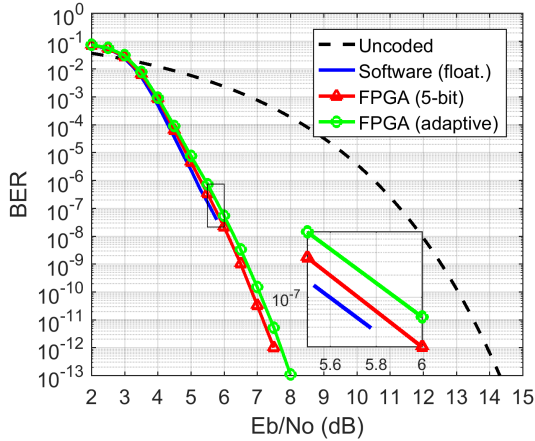}
	\caption{BER performance of (1024,854) polar code with SC decoder}
	\label{fig:sc_ber}
\end{figure}
% ----------------------------------------
\subsection{FPGA resource utilization} \label{sec:fpga_performance}
% ----------------------------------------
FPGA resource utilization is shown in Table \ref{table:resource_utilization_fpga}. 2-core implementation has the lowest total power and resource usage in terms of CLB, LUT and FF. The power per core is the lowest for 8-core implementation.
\begin{table}[ht]
\centering
\caption{FPGA resource utilization}
\label{table:resource_utilization_fpga}
\begin{tabular}{lccc}
\hline
\textbf{Decoder cores}       &  2-core & 4-core & 8-core \\ 
\textbf{Pipeline depth}      & 25        & 25   & 12          \\ 
\textbf{Core freq. (MHz)}     & 50       & 30       & 30        \\
\textbf{Throughput (Gb/s)}   &  85.4      & 102.5        & 204.9       \\ \hline
\textbf{Total CLB}           & 19568  & 38548  & 75683 \\ 
\textbf{Total LUT}           & 109954 & 219894 & 376805  \\ 
\textbf{Total FF}            & 76646  & 152283 & 276077 \\ 
\textbf{Total BRAM}          & 272    & 544    & 228 \\ 
\textbf{Total power} (mW)    & 484    & 644    & 1024\\ \hline
\textbf{CLB per core}        & 12090  & 12657  & 11681 \\ 
\textbf{LUT per core}        & 52451  & 52445  & 47123 \\ 
\textbf{FF per core}         & 22486  & 22485  & 19041 \\ 
\textbf{BRAM per core}       & 136    & 136    & 28.5 \\ 
\textbf{Power per core} (mW)& 211    & 129    & 99 \\  \hline
\end{tabular}
\end{table}
% ---------------------------------------- 
\section{ASIC Implementation} \label{sec:asic_implemetation}
% ----------------------------------------

This section presents ASIC implementation design flow information and results for each design. The information is as follows.
\begin{itemize}
\item The TSMC 28nm CMOS logic High Performance Computing (HPC) library (tcbn28hpcbwp12t30p140ssg0p72v125c) is used for RTL synthesis and backend implementation.
\item The regular voltage threshold (RVT) standard logic cells are used.
\item The setup and hold time of each design are verified for typical, worst-C, worst-RC, best-C and best-RC design corners. To achieve timing-clean results, a noticeable number of buffers and inverters have been added to the design. 
\item Power is estimated using vectorless power analysis at a constant 0.2 activity factor (default) for input signals.
\item The final implementation results are obtained at the end of a timing-clean P\&R.
\end{itemize}

% ----------------------------------------
\subsection{Synthesis} \label{sec:synthesis}
% ----------------------------------------
The synthesis results of MCSC decoders are shown in Table \ref{table_multicore_synthesis_results}. The results include a single instance of each MCSC decoder. The results show that as the number of decoder cores increase, the area-per-core reduces significantly due to efficient optimization of pipeline depth using R-RB scheme.
\begin{table}[ht]
\centering
\caption{Synthesis results of the multicore polar SC decoders}
\label{table_multicore_synthesis_results}
\begin{tabular}{lcccc}
\hline
\textbf{Num. of cores}                    & 1      & 2      & 4      & 8      \\ 
\textbf{Core clock freq.} (MHz)       & 1200   & 600    & 300    & 150    \\ 
\textbf{Core pipeline depth}             & 124    & 59     & 25     & 13     \\ 
\textbf{Num. of cells per core}             & 941171 & 727774 & 382681 & 199751 \\ 
\textbf{Total area} (mm$^2$)      & 2.60   & 3.72  & 3.64   & 3.60   \\ 
\textbf{Area per core} (mm$^2$)      & 2.60   & 1.86   & 0.91   & 0.45   \\ 
\textbf{Cell area per core} (mm$^2$)         & 2.25   & 1.44   & 0.70   & 0.35   \\ \hline
\end{tabular}
\end{table}

% ----------------------------------------
\subsection{Floorplan} \label{sec:floorplan}
% ----------------------------------------
The floorplans of the MCSC decoders are shown in Fig. \ref{fig:floorplan}. LLR input pins are located at top and decision outputs are located at bottom. The remaining control and clock pins are located at left. The 1-core design has 3.3 mm$^2$ square floorplan to maximize the area efficiency. Other designs have a rectangular floorplan with 1-to-2 ratio to achieve better combined multicore layout.

\begin{figure}[]
	\centering
	\includegraphics[scale=0.5]{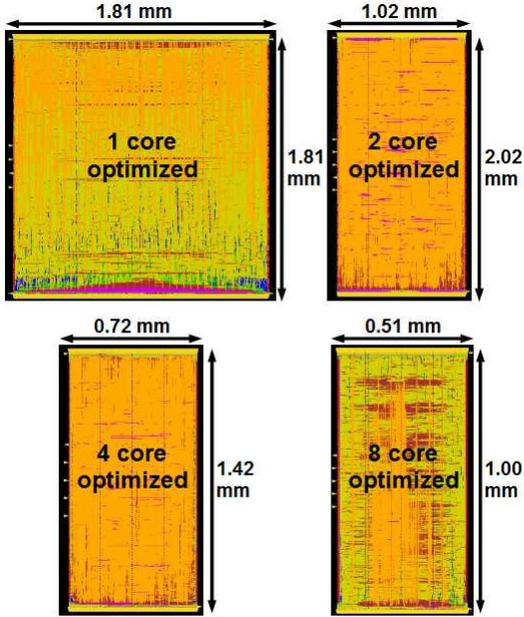}
	\caption{Floorplans of (1024, 854) MCSC decoders}
	\label{fig:floorplan}
\end{figure}

% ----------------------------------------
\subsection{ASIC implementation results} \label{sec:asic_results}
% ----------------------------------------
Post-P\&R results of the MCSC decoder candidates are shown in Table \ref{table_multicore_post_pr_results}. All implementations achieve 1 Tb/s throughput with low power dissipation and reasonable power density. The 8-core implementation is favorable in terms of low core pipeline depth and low core clock frequency. The 1-core implementation is favorable in terms of low total area and area efficiency. On the other hand, the 4-core implementation seems to be the optimal choice in terms of low total power, latency, energy efficiency and power density results. The 4-core implementation has 2.3 times lower power and better energy efficiency while achieving almost 2.8 times better power density than 1-core implementation.

\begin{table}[]
\centering
\caption{Post-P\&R results of (1024, 854) MCSC decoders}
\label{table_multicore_post_pr_results}
\begin{tabular}{lcccc} \hline
\textbf{Architecture} & 1-core & 2-core & 4-core & 8-core \\
\textbf{Core frequency} (MHz) & 1200 & 600 & 300 & \textbf{150} \\
\textbf{Throughput} (Gb/s) & \textbf{1025} & \textbf{1025} & \textbf{1025} & \textbf{1025} \\
\textbf{Core pipeline depth} & 124 & 59 & 25 & \textbf{13} \\
\textbf{Total area} (mm$^2$) & \textbf{3.21} & 4.00 & 3.92 & 3.84 \\
\textbf{Cell area} (mm$^2$) & \textbf{2.25} & 2.90 & 2.80 & 2.83 \\
\textbf{Area utilization} (\%) & 70.0 & 72.5 & 71.5 & \textbf{73.6} \\
\textbf{Total power} (mW) & 3646 & 2232 & \textbf{1593} & 1673 \\
\textbf{Latency} (ns) & 105 & 102 & \textbf{90} & 100 \\
\textbf{Energy eff.} (pJ/b) & 3.56 & 2.18 & \textbf{1.55} & 1.63 \\
\textbf{Power dens.} (mW/mm$^2$) & 1134 & 558 & \textbf{407} & 436 \\
\textbf{Area eff.} (Gb/s/mm$^2$) & \textbf{319} & 256 & 262 & 267 \\ \hline
\end{tabular}
\end{table}

% ----------------------------------------
\subsection{Comparison with state-of-the-art} \label{SoAcomparison}
% ----------------------------------------
A comparison of 4-core MCSC decoder with other high-throughput polar SC decoders is shown in Table \ref{table:comparison}. The results are scaled to 16nm and 0.8 V for a fair comparison. The proposed MCSC decoder has more than 2.2 times better energy efficiency and 8\% better area efficiency that the other best singlecore SC decoder implementation. 

\begin{table}[ht]
\centering
\caption{Comparison with other high-throughput polar SC decoders}
\label{table:comparison}
\begin{tabular}{lccc} \hline
\textbf{Implementation}           & This work & \cite{Gross2017}$^\dagger$ & \cite{Sural2020} \\
\textbf{Architecture}  & 4-core & 1-core & 1-core \\
\textbf{ASIC technology}                           & 28nm                                                         & 28nm                         & 16nm                     \\
\textbf{Block length }                              & 1024                                                         & 1024                         & 1024                     \\
\textbf{Code rate }                                 & 0.83                                                         & 0.5                          & 0.83                     \\
\textbf{Supply voltage} (V)                         & 0.8                                                          & 1.0                          & 0.8                      \\ \hline
\textbf{Coded throughput} (Gb/s)                    & 1229                                                         & 1275                         & 1229                     \\
\textbf{Frequency }(MHz)                            & 300                                                          & 1245                         & 1200                     \\
\textbf{Area} (mm$^2$)                                 & 3.92                                                         & 4.63                         & 0.79                     \\
\textbf{Power} (mW)                                 & 1593                                                         & 8793                         & 1167                     \\ \hline
\multicolumn{4}{c}{Scaled to 16nm and 0.8V using the common scaling factors in \cite{Wong2010}.}                                                                                                                         \\ \hline
\textbf{Coded throughput} (Gb/s)                    & 2150                                                         & \textbf{2231}                         & 1229                     \\
\textbf{Core frequency }(MHz)                            & 525                                                         & 2179                         & 1200                     \\
\textbf{Latency }($\mu$s)                               & \textbf{0.05}                                                         & 0.17                         & \textbf{0.05}                     \\
\textbf{Area }(mm$^2$)                                 & 1.28                                                         & 1.51                         & \textbf{0.79}                     \\
\textbf{Area eff. }(Gb/s/mm$^2$)                       & \textbf{1681}                                                         & 1477                         & 1554                     \\
\textbf{Power }(mW)                                 & \textbf{910}                                                          & 3216                         & 1167                     \\
\textbf{Power density }(mW/mm$^2$) & \textbf{712}                                                          & 2128                         & 1473                     \\
\textbf{Energy eff. }(pJ/b)     & \textbf{0.42}                                                        & 1.44                         & 0.95                 \\   \hline
\multicolumn{4}{l}{$^\dagger$ Synthesis results are given.}\\
\end{tabular}
\end{table}

% ----------------------------------------
\section{Conclusion} \label{sec:conclusion}
% ----------------------------------------

This paper proposes a multicore polar SC decoder (MCSC) architecture to achieve Tb/s throughput. Post-P\&R results of the 1-core, 2-core, 4-core and 8-core MCSC candidate decoders are analyzed and compared to identify the most efficient implementation. The results show that the 4-core MCSC implementation has the lowest latency, total power, power density and the best energy efficiency results. 4-core MCSC decoder achieves 1 Tb/s throughput on 3.92 mm$^2$ area with 1.55 pJ/bit energy efficiency. The comparison with state-of-the-art shows that 4-core MCSC decoder has more than 2.2 times better energy efficiency and 2 times better power density with respect to next best Tb/s polar SC decoder.

% ----------------------------------------
\section*{Acknowledgment}
% ----------------------------------------
The work is an extension of work done in EPIC project funded by the European Union's Horizon 2020 research and innovation program under grant agreement No 760150.

\FloatBarrier
\bibliography{references}{}
\bibliographystyle{ieeetr}
\end{NoHyper}
\end{document}